\documentclass[amsfonts, amssymb, amsmath, twocolumn, showkeys, nofootinbib, reprint,floatfix, prl, superscriptaddress]{revtex4-2}

\usepackage[colorinlistoftodos, color=green!40,prependcaption]{todonotes}
\usepackage{graphicx}
\usepackage{dcolumn}
\usepackage{bm}
\usepackage{float}
\usepackage{gensymb}
\usepackage{xcolor}
\usepackage{amsmath}
\usepackage{amssymb}
\usepackage{units}
\usepackage[english]{babel}
\usepackage{placeins}
\usepackage{lipsum}
\usepackage{gensymb}

\begin{document}

\title{Spin-carrier coupling induced ferromagnetism and giant resistivity peak in EuCd$_2$P$_2$}

\author{V. Sunko}
\thanks{V. S. and Y. S. contributed equally to this work.}
\affiliation {Department of Physics, University of California, Berkeley, California 94720, USA}
\affiliation {Materials Science Division, Lawrence Berkeley National Laboratory, Berkeley, California 94720, USA}

\author{Y. Sun}
\thanks{V. S. and Y. S. contributed equally to this work.}
\affiliation {Department of Physics, University of California, Berkeley, California 94720, USA}
\affiliation {Materials Science Division, Lawrence Berkeley National Laboratory, Berkeley, California 94720, USA}

\author{M. Vranas}
\affiliation {Department of Physics, University of California, San Diego, California 92093, USA}

\author{C. C. Homes}
\affiliation {NSLS II, Brookhaven National Laboratory, Upton, New York 11973, USA}

\author{C. Lee}
\author{E. Donoway}
\affiliation {Materials Science Division, Lawrence Berkeley National Laboratory, Berkeley, California 94720, USA}

\author{Z.-C. Wang }
\author{S. Balguri}
\author{M. B. Mahendru}
\affiliation {Department of Physics, Boston College, Chestnut Hill, Massachusetts 02467, USA}

\author{A. Ruiz}
\author{B. Gunn}
\author{R. Basak}
\affiliation {Department of Physics, University of California, San Diego, California 92093, USA}

\author{E. Schierle}
\author{E. Weschke}
\affiliation {Helmholtz-Zentrum Berlin für Materialien und Energie, Albert-Einstein-Straße 15, 12489 Berlin, Germany}

\author{F. Tafti}
\affiliation {Department of Physics, Boston College, Chestnut Hill, Massachusetts 02467, USA}

\author{A. Frano}
\affiliation {Department of Physics, University of California, San Diego, California 92093, USA}

\author{J. Orenstein}
\affiliation {Department of Physics, University of California, Berkeley, California 94720, USA}
\affiliation {Materials Science Division, Lawrence Berkeley National Laboratory, Berkeley, California 94720, USA}

\begin{abstract}
EuCd$_2$P$_2$ is notable for its unconventional transport: upon cooling the metallic resistivity changes slope and begins to increase, ultimately 100-fold, before returning to its metallic value. Surprisingly, this giant peak occurs at \unit[18]{K}, well above the N\'{e}el temperature ($T_N$) of \unit[11.5]{K}. Using a suite of sensitive probes of magnetism, including resonant x-ray scattering and magneto-optical polarimetry, we have discovered that ferromagnetic order onsets above $T_N$ in the temperature range of the resistivity peak.  The observation of inverted hysteresis in this regime shows that ferromagnetism is promoted by coupling of localized spins and itinerant carriers. The resulting carrier localization is confirmed by optical conductivity measurements. \end{abstract}

\maketitle

First glimpses of phase transitions in new materials are often gained through measurement of the temperature-dependent resistivity, $\rho(T)$. At a magnetic transition $\rho(T)$ generically exhibits a change in slope, with $d\rho/dT$ proportional to the heat capacity anomaly. This behavior is understood within the Fisher-Langer theory, in which enhanced quasiparticle scattering appears as a consequence of critical fluctuations \cite{fisher_resistive_1968}. However, some metallic magnets exhibit features in $\rho(T)$ more pronounced than a mere change in slope. For example, resistivity peaks have been reported recently near the N\'{e}el temperature $T_N$ in a number of antiferromagnetic (AFM) europium compounds with the general formula EuM$_2$C$_2$ (M=In,Cd, C=Sb, As, P) \cite{ma_spin_2019,xu_unconventional_2021-1,rahn_coupling_2018,wang_colossal_2021,su_magnetic_2020,zhang_-plane_2020}. In these materials magnetism is localized in Eu$^{2+}$ layers, which are separated by low-carrier density itinerant M$_2$C$_2$ blocks. They are notable because density-functional theory predicts that their low-energy electronic structure and topology are dramatically altered by the nature of the magnetic order \cite{xu_higher-order_2019,hua_dirac_2018,soh_ideal_2019,wang_single_2019, soh_magnetic_2018,gui_new_2019,riberolles_magnetic_2021, taddei_single_2022},
and that the different magnetic states are close in energy \cite{krishna_first-principles_2018}, and are therefore experimentally accessible \cite{ma_spin_2019,jo_manipulating_2020,jo_visualizing_2021, gati_pressure-induced_2021}. This combination offers the exciting prospect of creating electronic states hypersensitive to external stimuli. 

\begin{figure}[t] 
\centering
\includegraphics[width=0.95\columnwidth]{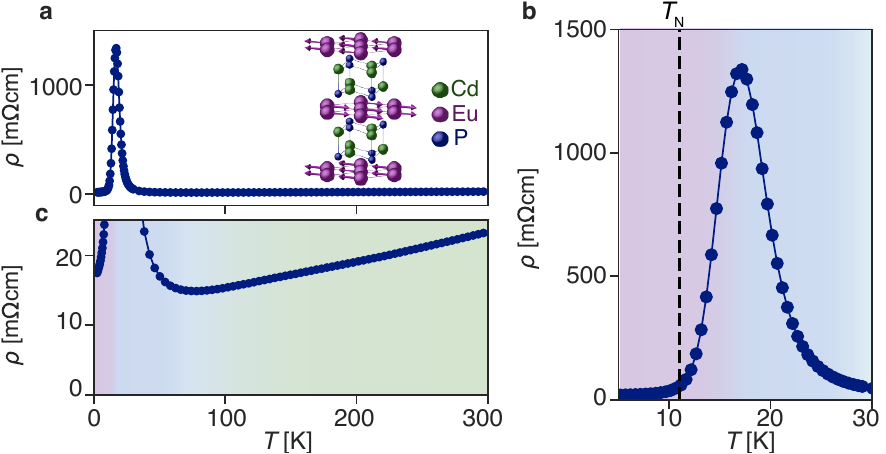}
\caption{(a,b) The resistivity of EuCd$_2$P$_2$ (structure in the inset) shows a pronounced peak at $\unit[18]{K}$. (c) Above $\unit[75]{K}$ it exhibits metallic behavior. Data are from ref.~\cite{wang_colossal_2021}.}
\label{fig:resisitivity}
\end{figure}

The resistivity of EuCd$_2$P$_2$, the subject of this study, stands out as an extreme example of unconventional temperature dependence, quantitatively and qualitatively different from other systems exhibiting a resistivity peak \cite{wang_colossal_2021}. As shown in Fig.~\ref{fig:resisitivity}, the metallic high-temperature resistivity (green in Fig.~\ref{fig:resisitivity}c) undergoes a hundredfold increase with decreasing temperature (blue) and subsequently returns to metallic values (purple). Both the rise and fall of the resistance take place well above the N\'{e}el temperature of $\unit[11.5]{K}$, as determined from the heat capacity measurements. The peak value of $\rho$, which is found at $\unit[18]{K}$, is suppressed by modest magnetic fields, yielding a giant negative magnetoresistance \cite{wang_colossal_2021}. 

Here we use a powerful combination of bespoke magneto-optical techniques and x-ray scattering to show that the anomalous resistivity peak arises from previously unreported time-reversal breaking that takes place above the N\'{e}el temperature. Our results indicate that AFM order is preceded by the formation of ferromagnetic clusters driven by the interaction of localized spin and itinerant charge degrees of freedom. The three resistivity regimes depicted in Fig.~\ref{fig:resisitivity} are traced to temperature-dependent crossovers from independent fluctuating clusters to the onset of ferromagnetic order.

We begin by describing our spatially-resolved symmetry-sensitive optical measurements, in which we measure the change of the angle of linear polarization ($d\phi$) as a function of sample orientation. Since we cannot physically rotate the sample fixed to piezoelectric scanning stages, we access the same information by rotating the incoming light polarization ($\phi$). Once the experimental setup is carefully aligned to preserve the polarization state (SI, Sec.~I), the change of polarization upon reflection is given by:
\begin{equation}
\label{eq:dphi}
d\phi\sim A\left(T\right)\sin\left[2\left(\phi-\phi_0\left(T\right)\right)\right]+B\left(T\right),
\end{equation}
where $\phi_0\left(T\right)$ and $\phi_0\left(T\right)+\pi/2$ indicate the directions of the principal optical axes, and $A\left(T\right)$ and $B\left(T\right)$ are proportional to birefringence and polar magneto-optical Kerr effect (pMOKE). These effects arise from the breaking of rotational symmetry and out-of-plane magnetization, $M_z$, respectively. To enhance experimental sensitivity, we modulate $T$ at a frequency $f\approx\unit[2]{kHz}$ using a second laser beam as a heater, essentially measuring the temperature derivatives of $A$ and $B$, $dA$ and $dB$ \cite{little_three-state_2020, sun_mapping_2021}.

\begin{figure}[t]
\centering
\includegraphics[width=0.92\columnwidth]{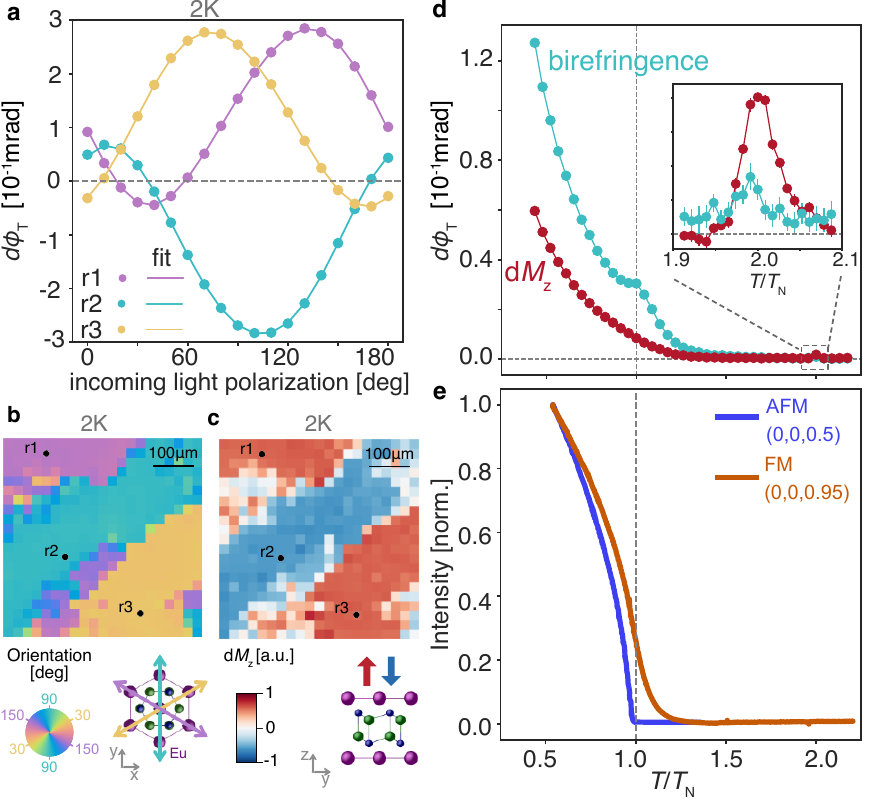}
\caption{(a) Thermally-modulated polarization rotation as a function of incident polarization and corresponding fits (Eq.~\ref{eq:dphi}), measured at three sample locations (probe: $\unit[20]{\mu W}$, $\unit[633]{nm}$; pump: $\unit[50]{\mu W}$, $\unit[780]{nm}$, modulated at $\unit[2345]{Hz}$).  (b, c) Maps of (b) principal axis orientation and (c) $dM_\text{z}$, extracted from the phase and offset of curves like  those in (a) (step size: $\unit[25]{\mu m}$, light spot diameter: $\unit[5]{\mu m}$). (d) Thermally-modulated birefringence amplitude (teal) and $dM_z$ (red) as a function of temperature ($T_N=\unit[11.5]{K}$). Both signals onset at $\sim2T_N$ (inset). (f) Normalized antiferromagnetic (blue; $(h,k,l)=(0, 0, 0.5)$) and ferromagnetic (orange; $(h,k,l)=(0, 0, 0.95)$, with structural scattering subtracted) REXS as a function of temperature. AFM scattering onsets sharply at $T_N$, in contrast to the gradual onset of ferromagnetism above $T_N$.}
\label{fig:Fig2_exp}
\end{figure}

This experiment at $\unit[2]{K}$ reveals a sinusoidal dependence of $d\phi$ on $\phi$ (Fig.~\ref{fig:Fig2_exp}a, Eq.~\ref{eq:dphi}), corresponding to a nonzero value of $dA$ and indicating broken rotational symmetry. Fig.~\ref{fig:Fig2_exp}b is a map of the spatial distribution of $\phi_0$, revealing three birefringent domains with $\unit[120]{\degree}$ between their principal axes (see Fig.~S1a of SI for histogram of the relative domain populations). This observation is consistent with spontaneous breaking of $C_3$ symmetry at the onset of previously reported type A-AFM order (i.e., alternating ferromagnetic layers with in-plane spins, ref.~\cite{wang_colossal_2021}) on a triangular lattice.

The first indication of unconventional order in EuCd$_2$P$_2$ is the observation of nonzero $dB$, indicated by the $\phi$-independent offset of the sinusoidal curves in Fig.~\ref{fig:Fig2_exp}b. As mentioned above, this offset demonstrates the existence of an out-of-plane component of magnetization, $M_z$; we therefore refer to it as $dM_z$ for clarity. The $dM_z$ map (Fig.~\ref{fig:Fig2_exp}c) shows that domains of $M_z$ and A-AFM order are highly correlated, suggesting coexistence and strong coupling between the two forms of order. In contrast, reflectivity is uniform across this sample region (see Figs.~S1(b-d)). 

The temperature dependence of $dA$ shown in Fig.~\ref{fig:Fig2_exp}d exhibits an additional surprising feature: the birefringence that was provisionally attributed to antiferromagnetism does not vanish at $T_N$. Rather, with increasing temperature there is a discontinuous change in slope at $T_N$ followed by a gradual decrease in amplitude, showing that $C_3$ symmetry remains broken above $T_N$. $dM_z$ also remains nonzero above $T_N$, with the polar Kerr signal changing smoothly through the transition. Zooming in on the weak feature at $\sim 2T_N$, we see peaks in both temperature-modulated birefringence and pMOKE (Fig.~\ref{fig:Fig2_exp}d, inset), showing that the two order parameters onset simultaneously, and are likely therefore coupled. 

To help identify the origin of rotational symmetry breaking above $T_N$, we employed resonant elastic x-ray scattering (REXS), with photon energy tuned to the Eu M$_5$ edge ($\unit[1127.5]{eV}$). The amplitude of the (0 0 0.5) diffraction peak, corresponding to A-AFM order, drops sharply at $T_N=\unit[11.5]{K}$, ruling out A-AFM order as the source of $C_3$ breaking above $T_N$ (Fig.~\ref{fig:Fig2_exp}e). The ferromagnetic (FM) peak at (0 0 1) is obscured by the structural one; however a fortunate matrix element suppression of the structural Thomson scattering at (0 0 0.95) provides a window to the FM order (for details see SI, Sec.~II). In contrast with the AFM onset, the FM order decreases smoothly through $T_N$ (Fig.~\ref{fig:Fig2_exp}f), mirroring the $dM_z(T)$ extracted from the pMOKE signal (Fig.~\ref{fig:Fig2_exp}d). We note that our measurements indicate static FM order, in contrast to short-range fluctuations reported in the sibling compound EuCd$_2$As$_2$ \cite{soh_resonant_2020}.

Having confirmed the existence of ferromagnetic order for $T>T_N$, we turn to the origin of rotational symmetry breaking above the N\'{e}el temperature. To that end, we have developed a method, based on the linear magneto-optic effect \cite{kharchenko_light_1978,kharchenko_lowering_1978,kharchenko_linear_1981,eremenko_linear_1985,eremenko_magneto-optics_1987,eremenko_birefringence_1980, eremenko_magneto-optical_1992,kharchenko_linear_1994}, to isolate rotational symmetry breaking associated with order parameters that break time-reversal symmetry, as well. Specifically, we measure the modulation of birefringence that is linear in an applied magnetic field; we refer to this effect as linear magneto-birefringence (LMB), and parameterize it by the LMB tensor,
\begin{equation} \label{eq:reflectivity}
\tensor{\delta r}=\left(\begin{array}{cc}
\beta & \gamma \\
\gamma &-\beta 
\end{array}\right)H_z,
\end{equation} 
where $\tensor{\delta r}$ is the change of reflectance, $H_z$ is an applied field, and $\left(1,0\right)$ and $\left(0,1\right)$ correspond to the principal axes at $H_k=0$. As a consequence of Onsager's relation,  the LMB tensor vanishes in time-reversal (TR) invariant systems (see SI, Sec.~III for a derivation), making it a sensitive probe of TR symmetry breaking.

We measure the components of $\tensor{\delta r}$ by detecting the change in reflectivity at the fundamental frequency of an oscillating $H_z$, applied by a small coil \cite{lee_observation_2022}. After careful calibration of the setup (SI, Sec.~I.C), the change in polarization on reflection that is synchronous with the applied field, $d\phi\sim AH_z\sin\left[2\left(\phi+\phi_C\right)\right]$,
yields the elements of the LMB tensor through the relations,
\begin{equation} \label{eq:amplitudeCoil} 
A^{2}=\beta^{2}+\gamma^{2},\hspace{0.3cm} \phi_{C}=\phi_{0}+\frac{1}{2}\arctan\frac{\gamma}{\beta},
\end{equation} 
where $\phi_0$ is the principal axes orientation at $H_z=0$. 

\begin{figure}[t]
\centering
\includegraphics[width=0.95\columnwidth]{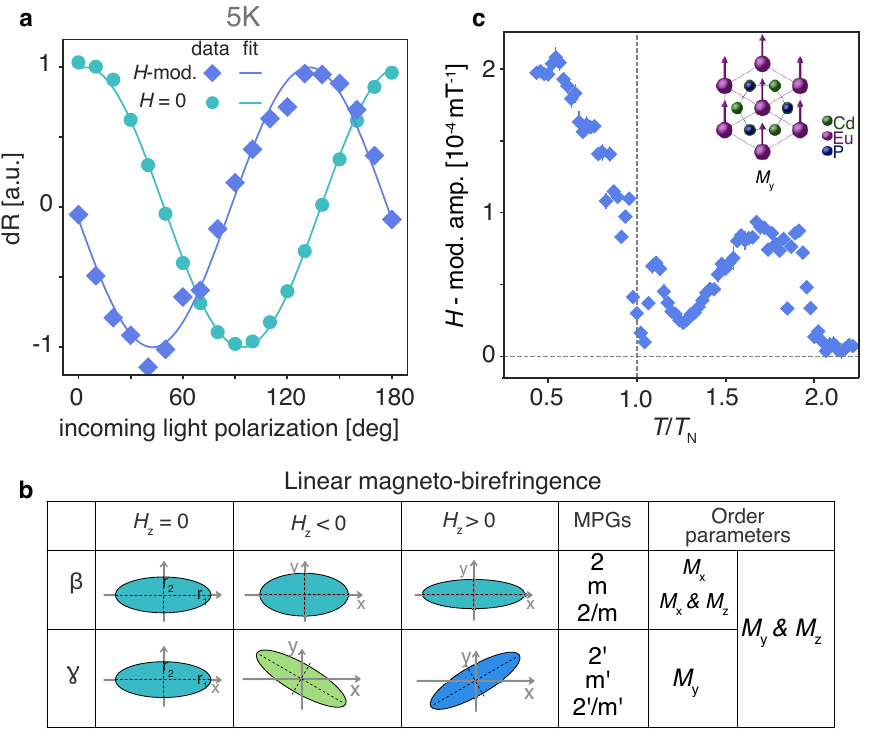}
\caption{(a) The phase difference between the $H_z$-linear birefringence (diamonds) and the $H=0$ birefringence (circles) proves $\gamma\neq0$ (Eq.~\ref{eq:amplitudeCoil}), and therefore $M_y\neq0$ (inset of (c)). For clarity the curves are normalized by the respective amplitudes, and constant offsets are subtracted. (b) Diagonal ($\beta$) and off-diagonal ($\gamma$) LMB: the orientation and length of the ellipse axes represent the orientation of the optic principal axes and the reflectivity along them, respectively. Magnetic point groups (MPGs) and order parameters compatible with the two effects are listed. (c) The amplitude of LMB as a function of temperature ($T_N=\unit[11.5]{K}$) reveals the onset of TR - and $C_3$- broken phase at $\sim2T_N$. Probe beam: $\unit[20]{\mu W}$, $\unit[633]{nm}$; field amplitude: $\unit[0.2]{mT}$, modulated at $\unit[500]{Hz}$.}
\label{fig:Fig3_v1}
\end{figure}

Fig.~\ref{fig:Fig3_v1}a shows the existence of LMB in EuCd$_2$P$_2$ at $\unit[5]{K}$. The diamonds and circles compare the change in polarization induced by field and temperature modulation, respectively, and demonstrate a phase shift of $\unit[52]{\degree}$ between them. The observation of a phase shift that is not equal to either $\unit[0]{\degree}$ or $\unit[45]{\degree}$ proves that both $\beta$ and $\gamma$ are nonzero (Eq.~\ref{eq:amplitudeCoil}). A symmetry analysis of the LMB tensor in EuCd$_2$P$_2$, summarized in Fig.~\ref{fig:Fig3_v1}b and discussed in detail in Sec.~III of SI, shows that a nonzero $\gamma$ requires a  $M_y$, revealing this component of FM order at $\unit[5]{K}$.

The temperature dependence of the field-modulated amplitude, plotted in Fig.~\ref{fig:Fig3_v1}c, shows that LMB remains finite above $T_N$, identifying $M_y$ as the origin of birefringence above the AFM transition. It finally vanishes at $\sim2T_N$, coincident with the disappearance of $M_z$ (Fig.~\ref{fig:Fig2_exp}d). We note that $M_y$ and $M_z$ correspond to distinct magnetic point groups, so one is not a natural consequence of the other, as in `weak ferromagnetism' \cite{dzyaloshinsky_thermodynamic_1958}, for example. Their coexistence indicates coupling via a high-order magnetocrystalline anisotropy (MCA), which typically arises from a combination of spin-orbit coupling and the crystal field \cite{coey_anisotropy_2021}. However, the ground state of Eu$^{2+}$ carries no angular momentum, so spin-orbit coupling is expected to be negligible and the anisotropy cannot be understood by considering only Eu spins. Instead, the MCA can stem from the coupling of Eu to carriers, as was previously demonstrated in other magnets containing the $L=0$ Eu$^{2+}$ or Gd$^{3+}$ ions \cite{colarieti-tosti_origin_2003,abdelouahed_magnetic_2009,blanco-rey_effect_2022}.

\begin{figure}[t]
\centering
\includegraphics[width=0.87\columnwidth]{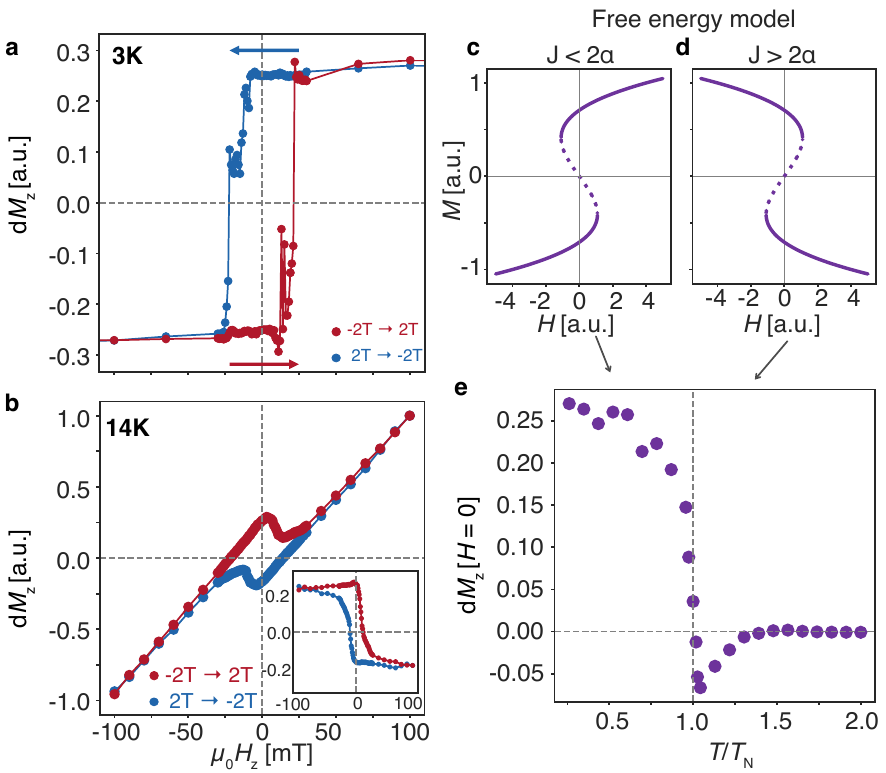}
\caption{$dM_\text{z}$ as a function of $H_z$ for increasing (red) and decreasing (blue) field at (a) \unit[3]{K} and (b) \unit[14]{K}, revealing the opposite sense of hysteresis at the two temperatures. Removing the linear background in (b) shows the inverted hysteresis (inset). (c,d) The real solutions of Eq.~\ref{eq:cubic}, for (c) $J<2\alpha$ and (d) $J>2\alpha$. The full and dotted lines correspond to local minima and saddle points of the free energy, respectively. (e) The temperature dependence of the magnetization at zero field, defined as half the difference between the $dM_z\left(H=0\right)$, measured while increasing and decreasing the field. }
\label{fig:Fig4_v1}
\end{figure}

Confirmation that ferromagnetism in EuCd$_2$P$_2$ results from the coupling between Eu spins and conduction electrons comes from measurements of $dM_z$ as a function of applied dc magnetic field. Fig.~\ref{fig:Fig4_v1}a follows the magnetization through a field sweep at $\unit[3]{K}$, showing a hysteresis loop that confirms FM order. The same field sweep conducted at $\unit[14]{K}$ (Fig.~\ref{fig:Fig4_v1}b) is also hysteretic, but looks very different in two respects: the loop proceeds clockwise rather than counterclockwise, and is superposed on a linear background. Removing the background (inset) emphasizes the hysteresis is \textit{inverted}: the saturated magnetization is anti-parallel to the applied field direction!

Inverted hysteresis cannot arise from a single class of ferromagnetically coupled spins; instead, at least two \textit{antiferromagnetically} coupled subsystems are required. A minimum model for the observed behavior is described by the free energy,
\begin{equation}\label{eq:freeEnergy} 
F=\alpha_2m^{2}+\beta_2M^{2}+\beta_4 M^{4}+JmM-H(m+M),
\end{equation}
where $m$ and $M$ are the magnetization of carriers and Eu ions, respectively, $J$ is the antiferromagnetic coupling between them, and the parameters $\alpha_n$ and $\beta_n$ are the quadratic and quartic terms in the expansion in even powers of the magnetization. With $H=0$, $F$ has a minimum at nonzero magnetization, 
\begin{equation} \label{eq:magnetization} 
M^2=\frac{1}{4\beta_4}\left(\frac{J^2}{2\alpha_2}-2\beta_2\right),\hspace{0.3cm} m=-\frac{J}{2\alpha_2}M,
\end{equation}
for coupling strengths $J^2>4\beta_2\alpha_2$, showing that $J$ promotes spontaneous magnetization, even if the uncoupled Eu system would be paramagnetic ($\beta_2>0$). 

Minimizing the free energy with $H_z\neq 0$ yields (SI, Sec.~IV): 
\begin{equation} \label{eq:cubic}
M^3-\frac{M}{4\beta_4}\left(\frac{J^2}{2\alpha_2}-2\beta_2\right)-\frac{H}{4\beta_4}\left(1-\frac{J}{2\alpha_2}\right)=0.
\end{equation}
Eq.~\ref{eq:cubic} suggests that the effective magnetic field experienced by the Eu spin is opposite in sign to the applied field if $J$ is larger than $2\alpha_2$, yielding the inverted Eu hysteresis. Although the total magnetization, $M+m$, is parallel to the applied field, our experiment is dominantly sensitive to Eu magnetization, and therefore capable of observing its inverted hysteresis. The existence of both normal and inverted Eu hysteresis within this model is demonstrated in Figs.~\ref{fig:Fig4_v1}(c, d), which illustrate $M$ vs. $H$ for $J<2\alpha_2$ and $J>2\alpha_2$, respectively. Both regimes exhibit classic bistable behavior, with three solutions for $\partial F/\partial M=\partial F/\partial m=0$ at low applied field; two stable minima (solid lines) and one saddle point (dashed line). Mirrored trajectories of $M$ vs. $H_z$ in Figs.~\ref{fig:Fig4_v1}(c, d) correspond to normal and inverted hysteresis, respectively.

The temperature dependence of the zero-field magnetization (Fig.~\ref{fig:Fig4_v1}e), defined to be positive for normal hysteresis, shows that the transition from normal to inverted regimes occurs at $T_N$. The onset of the antiferromagnetic order therefore disrupts the delicate balance required for realization of the inverted hysteresis. Our model suggests that this is caused by a reduction in the effective spin-carrier interaction, $J/\alpha_2$, as expected from the decrease in the spin-polarization of the electrons due to their coupling to the AFM order.

The observations of FM order and inverted hysteresis above $T_N$ offer compelling evidence for the role of spin-carrier coupling in determining the magnetic properties of EuCd$_2$P$_2$. We now turn to the role of this coupling in its remarkable resistivity. To help address this question we augmented our magneto-optical and x-ray probes with measurements of the frequency- and temperature- dependent conductivity, $\sigma_1\left(\omega,T\right)$, obtained by Kramers-Kronig analysis of broadband reflectivity. 

\begin{figure}[t]
\centering
\includegraphics[width=0.87\columnwidth]{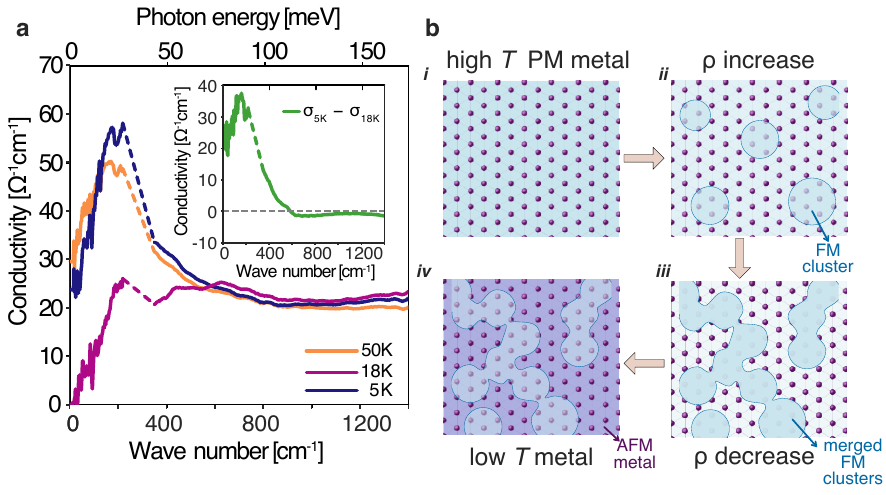}
\caption{(a) Optical conductivity at $\unit[50]{K}$, $\unit[18]{K}$ and $\unit[5]{K}$, showing electron localization at $\unit[18]{K}$. The inset is the difference between the $\unit[5]{K}$ and the $\unit[18]{K}$ data, showing the conduction electron spectral weight. The dashed lines are interpolations across a region where the measurement is dominated by the spectral weight of an optical phonon. (b) Schematic temperature evolution of the coupled Eu-electron system.}
\label{fig:Fig5_v1}
\end{figure}

Fig.~\ref{fig:Fig5_v1}a shows optical conductivity spectra measured at the resistivity peak ($\unit[18]{K}$), together with representative spectra measured in the metallic regime at temperatures above ($\unit[50]{K}$) and below ($\unit[5]{K}$). The contrast in the spectra between the high and low conductivity regimes provides a clear picture of the dynamics that underline the giant resistance peak. While the optical conductivity at $\unit[5]{K}$ and $\unit[50]{K}$ is remarkably similar, $\sigma_1(\omega, \unit[18]{K})$ is clearly suppressed at low frequencies. Subtracting the spectrum measured at $\unit[18]{K}$ from that measured at $\unit[5]{K}$ reveals a Drude-like peak (Fig.~\ref{fig:Fig5_v1}a, inset). We attribute the metallic conductivity observed in the high and low temperature regimes to this component of $\sigma_1(\omega)$; the resistivity peak at $\unit[18]{K}$ then reflects its vanishing. We note that standard analysis of the optical conductivity and Hall effect (SI, Sec.~V.) yields a carrier concentration $n_e=\unit[0.6\times10^{19}]{cm^{-3}}$, scattering rate $\Gamma=9\times10^{13}$s$^{-1}$, and a mass of $1.5m_e$, where $m_e$ is the free electron mass. The conductivity spectra measured as $T$ passes through the peak in resistivity confirm that the increase in $\rho$ originates from a dramatic decrease in $n_e$ rather than an increase in $\Gamma$. Optical conductivity therefore offers unambiguous evidence of electron localization. 

Peaks in resistivity have been observed in several metallic magnets, but a unifying theoretical explanation for this unconventional behavior is lacking \cite{timm_disorder-induced_2005,wang_resistivity_2016,flebus_magnetoresistance_2021, heinrich_topological_2022}. EuCd$_2$P$_2$ is an outlier in both the giant amplitude and location in temperature of the maximum in $\rho(T)$.  Nonetheless, our combination of optical polarimetry, x-ray, and far-infrared measurements points to spontaneous phase separation, similar to that observed in the \emph{ferromagnetic} metal EuB$_6$ \cite{pohlit_evidence_2018, das_magnetically_2012, brooks_magnetic_2004}. In this scenario (Fig.~\ref{fig:Fig5_v1}b) high-temperature metallic behavior is interrupted by phase separation: electrons localize into spin-polarized clusters, within which spins align due to the spin-carrier coupling ($J$ in Eq.~\ref{eq:freeEnergy}). The localization (Fig.~\ref{fig:Fig5_v1}a) causes the resistivity to increase until the clusters merge and form percolating conducting paths, slowing down the resistivity increase and eventually causing its decrease. The merged clusters develop a net ferromagnetic moment ($M_y$, $M_z$), detected by our suite of probes (REXS, pMOKE and LMB). In particular, the spontaneous phase separation is seen in the behavior of the magnetization vs. $H_z$ in this regime (Fig.~\ref{fig:Fig4_v1}b): paramagnetic Eu spins contribute the linear background, while the inverted hysteresis arises in the carrier-mediated ferromagnetic regions. 

Our finding that the same process, i.e. the spontaneous formation and percolation of ferromagnetic clusters, is responsible for resistivity in both EuCd$_2$P$_2$ and EuB$_6$, is surprising. The two compounds exhibit different magnetic order, and drastically different resistivity anomalies: the 100-fold increase of $\rho$ in the AFM EuCd$_2$P$_2$ \textit{versus} the 20\% increase in the FM EuB$_6$. We suggest that the low carrier density and the frustration resulting from near degeneracy of ferro- and antiferromagnetic states in EuCd$_2$P$_2$ are the key ingredients yielding the enhanced effect. These ingredients are then principles that can guide the design of other systems in which transport displays hypersensitivity to external control parameters such as temperature and electromagnetic fields. Furthermore, our work motivates the development of a unified theoretical description of resistivity in a wide range of systems with strong carrier-spin interactions, and offers ways to experimentally constrain such models.

\begin{acknowledgments} 
 
We thank Elbio Dagotto, Andrew Mackenzie, Chunxiao Liu, Marc Vila Tusell, Thomas Scaffidi and Ehud Altman for useful discussions. Optical measurements were performed at the Lawrence Berkeley Laboratory as part of the Quantum Materials program, Director, Office of Science, Office of Basic Energy Sciences, Materials Sciences and Engineering Division, of the U.S. Department of Energy under Contract No. DE-AC02-05CH11231. V.S. is supported by the Miller Institute for Basic Research in Science, UC Berkeley. J.O and Y.S received support from the Gordon and Betty Moore Foundation's EPiQS Initiative through Grant GBMF4537 to J.O. at UC Berkeley. F. T. and S. B. acknowledge funding from the Schiller Institute Grant for Exploratory Collaborative Scholarship (SIGECS). Work at Brookhaven National Laboratory was supported by the Office of Science, U.S. Department of Energy under Contract No. DE-SC0012704

\end{acknowledgments}

\bibstyle{apsrev4-1}

\bibliography{EuCd2P2_main}

\end{document}


\title{Spin-carrier coupling induced ferromagnetism and giant resistivity peak in EuCd$_2$P$_2$: Supplementary Information}

\author{V. Sunko}
\thanks{V. S. and Y. S. contributed equally to this work.}
\affiliation {Department of Physics, University of California, Berkeley, California 94720, USA}
\affiliation {Materials Science Division, Lawrence Berkeley National Laboratory, Berkeley, California 94720, USA}

\author{Y. Sun}
\thanks{V. S. and Y. S. contributed equally to this work.}
\affiliation {Department of Physics, University of California, Berkeley, California 94720, USA}
\affiliation {Materials Science Division, Lawrence Berkeley National Laboratory, Berkeley, California 94720, USA}

\author{M. Vranas}
\affiliation {Department of Physics, University of California, San Diego, California 92093, USA}

\author{C. C. Homes}
\affiliation {CMPMS, Brookhaven National Laboratory, Upton, New York 11973, USA}

\author{C. Lee}
\author{E. Donoway}
\affiliation {Materials Science Division, Lawrence Berkeley National Laboratory, Berkeley, California 94720, USA}

\author{Z.-C. Wang }
\author{S. Balguri}
\author{M. B. Mahendru}
\affiliation {Department of Physics, Boston College, Chestnut Hill, Massachusetts 02467, USA}

\author{A. Ruiz}
\author{B. Gunn}
\author{R. Basak}
\affiliation {Department of Physics, University of California, San Diego, California 92093, USA}

\author{E. Schierle}
\author{E. Weschke}
\affiliation {Helmholtz-Zentrum Berlin für Materialien und Energie, Albert-Einstein-Straße 15, 12489 Berlin, Germany}

\author{F. Tafti}
\affiliation {Department of Physics, Boston College, Chestnut Hill, Massachusetts 02467, USA}

\author{A. Frano}
\affiliation {Department of Physics, University of California, San Diego, California 92093, USA}

\author{J. W. Orenstein}
\affiliation {Department of Physics, University of California, Berkeley, California 94720, USA}
\affiliation {Materials Science Division, Lawrence Berkeley National Laboratory, Berkeley, California 94720, USA}

\maketitle


\begin{figure}[h] 
\centering
\includegraphics[width=0.9\columnwidth]{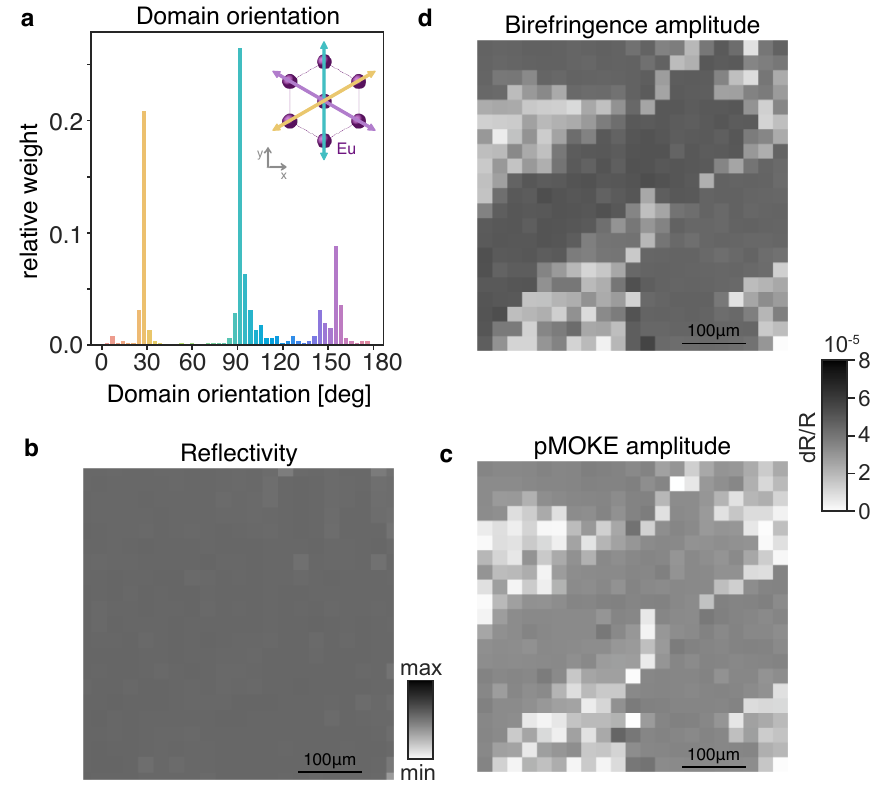}
\caption{ (a) Histogram of the relative domain populations in the map shown in Fig.~2b clearly shows three domains. (b) The reflectivity in the region mapped out in Fig.~2(b,c) of the main text is uniform.   The amplitude of (c) birefringence and (d) pMOKE over the same region are highly correlated: both are uniform within the large domains seen in  Fig.~2(b,c) of the main text, and smaller at the edges of those domains. None of the variability is caused by trivial variations of surface quality, as evidenced by the uniform reflectivity (panel (b)).}
\label{fig:histogram}
\end{figure}

\section{Optical setup and mitigation of artifacts}

\subsection{Experimental Setup}
\begin{figure}[h] 
\centering
\includegraphics[width=1\columnwidth]{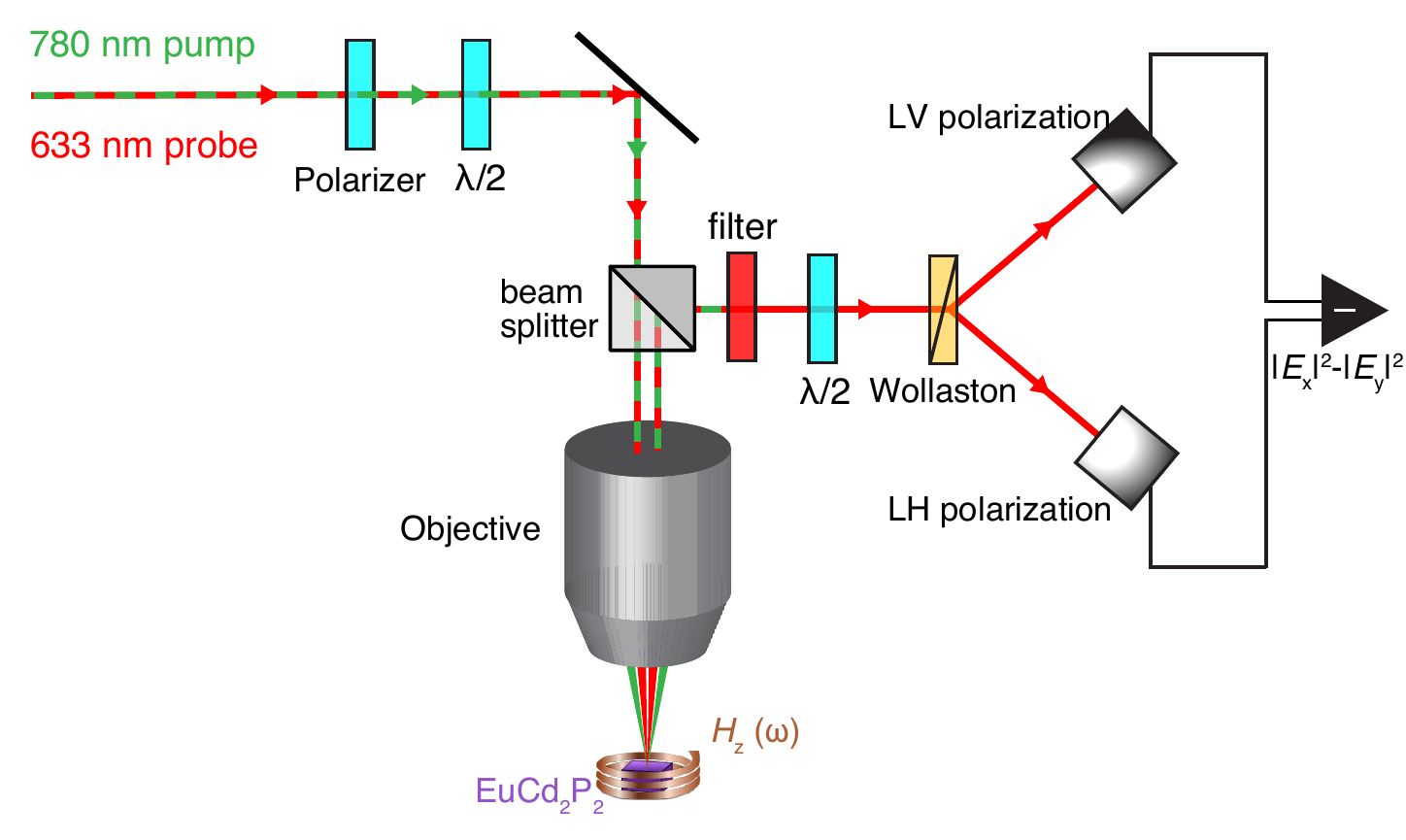}
\caption{Schematic of the optical setup used for measurements of the modulated polarization rotation as a function of incident polarization. In the thermally modulated configuration applied field is zero, while in the field-modulated configuration there is no pump beam.}
\label{fig:setup}
\end{figure}

A schematic of the optical setup is shown in Fig.~\ref{fig:setup}.  A polarizer is used to define the light polarization, followed by a half waveplate ($\lambda/2$), which sets the polarization angle, $\phi$. After reflecting off the sample, the polarization is rotated by $-\phi+45 \degree$; in other words, if the polarization state is not altered by the setup or the sample, the polarization is now an equal superposition of linear vertical (LV) and linear horizontal (LH) light. The beam is then split by the Wollaston prism into the two orthogonal linear polarizations, LV and LH,  and their difference is directly measured by a balanced detector. If the polarization state of the light remains unchanged by the sample, a zero is measured, while any measured signal indicates a change of polarization, making the setup very sensitive in detecting those changes. 

Unfortunately, changes of polarization can also be introduced by the birefringence of the setup, introducing artifacts. We largely mitigate this problem by performing temperature- and field- modulated experiments. For thermal modulation, we focus a second light beam (pump) at the same spot, using an optical chopper to modulate the pump beam at $\unit[]{kHz}$ frequencies. For field modulation, we place the sample inside of a coil, through which we pass an oscillating current. The experiment therefore becomes sensitive only to effects proportional to the modulation parameter, which the setup birefringence is not. Small cross-coupling terms can still occur if there is more than one optical constant proportional to the modulation parameter (symmetric reflectivity and birefringence, for example), as we discuss in  detail below. 

\subsection{Measured signal and mitigation of artifacts}

\begin{figure}[t] 
\centering
\includegraphics[width=0.9\columnwidth]{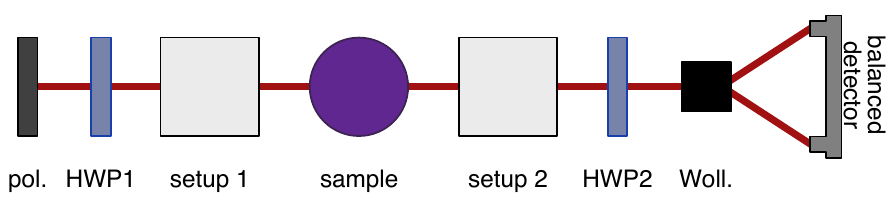}
\caption{Simplified schematic of the setup, indicating the terms included in the mathematical description: polarizer (pol), two half-wave plates (HWP), optical components before and after the sample (setup 1 and setup 2, respectively),  the Wollaston prism and the balanced detector. }
\label{fig:schemeSimple}
\end{figure}

In order to understand how the different terms influence the measured signal, we mathematically describe the experiment with the Jones matrix formalism: the light polarization state is represented by a vector in the $\left(\text{LV},\text{LH}\right)$ basis, and the effect of each optical element is encoded by a $2\times2$ matrix. Elements included in the mathematical description are shown in the simplified schematic (Fig.~\ref{fig:schemeSimple}). We treat separately cases with no modulation (for illustration purposes only - this configuration is not used in the experiments), with thermal modulation, and with magnetic field modulation.

The half-wave plate whose fast axis is rotated by $\theta$ with respect to the horizontal axis, rotates the polarization by $\phi=2\theta$, and is described by: 

\begin{equation}\label{eq:HWP}
    J_{hwp}\left(\theta\right)=\left(\begin{array}{cc}
\cos(2\theta) & \sin(2\theta)\\
\sin(2\theta) & -\cos(2\theta)
\end{array}\right).
\end{equation}
The ideal setup would not alter the polarization state, but any realistic one will; we therefore describe it as: 

\begin{equation}
    J_{exp}\left(\epsilon,\delta\right)=\left(\begin{array}{cc}
1 & 0\\
0 & e^{\imath\delta}(\epsilon+1),
\end{array}\right),
\end{equation}
where $\epsilon$ and $\delta$ mark the difference in amplitude and phase between the reflected vertical and horizontal light (in an ideal setup $\epsilon=\delta=0$). The effect of $\epsilon$, which we refer to as the setup birefringece, is to change the angle of linear polarization, unless the polarization is exactly aligned with one of the principal axes of the setup. The effect of $\delta$, the setup retardance, is to change the phase, making linear light elliptical.

The Jones matrix representing the sample is:

\begin{equation}
        J_{sam}\left(r,b,k, \phi_0\right)=R\left(-\phi_{0}\right)\left(
\begin{array}{cc}
 r+b & k \\
 -k & r-b \\
\end{array}
\right)R\left(\phi_{0}\right),
\end{equation}
where $r$ stands for the sample reflectivity, $b$ for the birefringence (difference in reflectivity between the LH and LV polarizations), and $k$ for the polar Kerr effect (difference in reflectivity between the left- and right-circularly polarized light caused by an out-of-plane magnetisation). $R\left(\phi_{0}\right)$ is the rotation matrix, allowing for the different orientations of the sample with respect to the lab coordinate system: 

\begin{equation}
    R\left(\phi_0\right)=\left(
\begin{array}{cc}
 \cos (\phi_0 ) & -\sin (\phi_0 ) \\
 \sin (\phi_0 ) & \cos (\phi_0 ) \\
\end{array}
\right)
\end{equation}
The state of  polarization of the light that traveled through the system is given by: 

\begin{equation}
\begin{split}
\MoveEqLeft
\left(\begin{array}{c}
E_{V}\\
E_{H}
\end{array}\right)=J_{hwp}\left(\phi/2+22.5\deg\right)\\ 
& \times J_{exp}\left(\epsilon_{2},\delta_{2}\right)
\times J_{sam}\left(r,b,k, \phi_0\right)\\
&\times J_{exp}\left(\epsilon_{1},\delta_{1}\right)\times J_{hwp}\left(\phi/2\right)\times\left(\begin{array}{c}
1\\
0
\end{array}\right),
\end{split}
\end{equation}
leading to the measured intensity: 

\begin{equation}
    I = |E_{V}|^2-|E_{H}|^2.
\end{equation}
In the experiment, we rotate the half-wave plates, leading to the polarization rotation ($\phi$); this polarization rotation is mimicking a sample rotation, while ensuring that the location of the light spot on the sample does not change. In the following, we discuss the dependence of the measured intensity on the polarization angle, $I\left(\phi\right)$, for the various experimental configurations. 

\subsection{No modulation}
\subsubsection{Ideal setup}
If the setup is ideal ($\epsilon=\delta=0$), the measured signal is equal to: 

\begin{equation}\label{eq:noMod_ideal}
   \frac{I\left(\phi\right)}{r^2}=  -\frac{2b}{r} \sin (2 (\phi +\phi_0 ))-\frac{2k}{r},
\end{equation}
where we neglect higher order terms in $b$. The observation that the birefringence ($b$) and the polar Kerr effect ($k$) are clearly separable because they manifest as the sinusoidal dependence on polarization and a polarization-independent offset, respectively, is the basis for our optical techniques (see Eq.~1 of the main text).  

\subsubsection{Setup birefringence and retardance}
The simple picture above is modified by the setup contributions. Even if the sample is not birefringent, a sinusoidal dependence can be measured. To first order in $\epsilon_1$ and $\epsilon_2$, it reads:

\begin{equation}\label{eq:setup}
   \frac{I_{setup}\left(\phi\right)}{r^2}= (\epsilon_1+\epsilon_2)\sin(2\phi) -\sin^{2}\left(\frac{\delta_1+\delta_2}{2}\right)\sin(4\phi).
\end{equation}
The effect of the setup birefringence ($\epsilon_1$ and $\epsilon_2$) and retardance ($\delta_1$ and $\delta_2$) can be distinguished, because of their different polarization dependence:  $\sin(2\phi)$ and $\sin(4\phi)$, respectively. Note that, unlike the signal arising from the sample, they do not depend on the sample orientation, $\phi_0$.

\begin{figure}[h] 
\centering
\includegraphics[width=0.7\columnwidth]{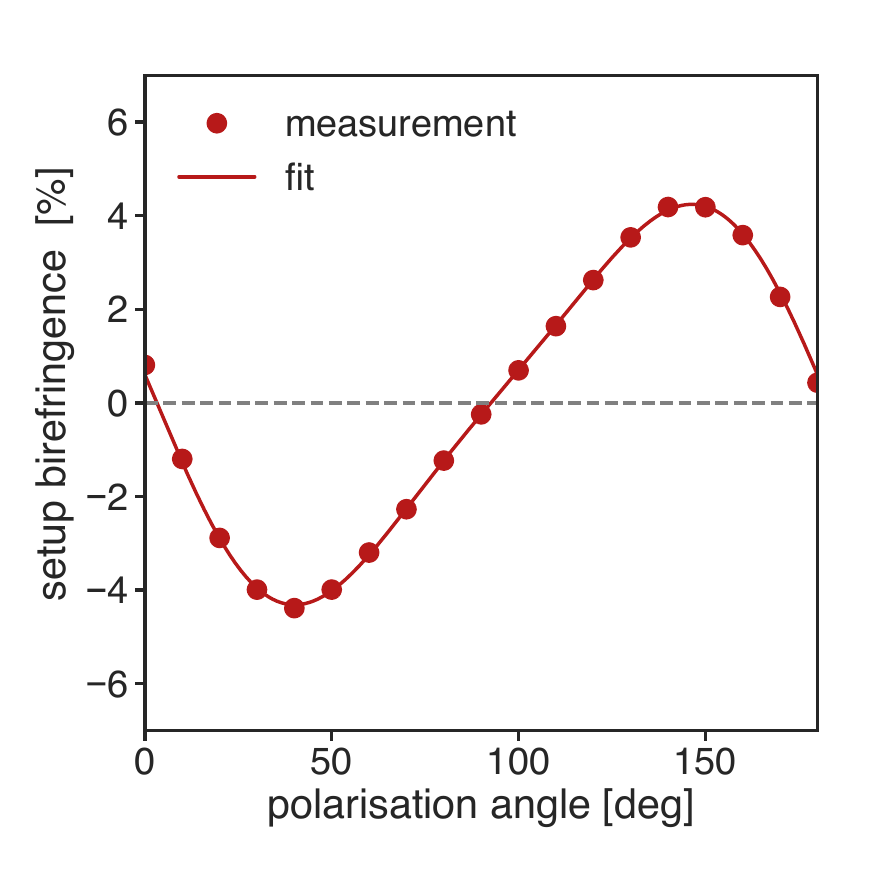}
\caption{polarization rotation experiment performed on a GaAs$\left(111\right)$ surface, revealing the setup contribution to a balanced measurement. The line is a fit to Eq.~\ref{eq:setup}, with $(\epsilon_1+\epsilon_2)=4.1\%$ and $\sin^{2}\left(\frac{\delta_1+\delta_2}{2}\right)=-0.7\%$}
\label{fig:GaAs_test}
\end{figure}

Setup contributions $\epsilon_{1,2}$ and $\delta_{1,2}$ can be minimized by orienting optical elements such that their contributions subtract, rather than add. The total effects of the setup can be measured by performing the usual polarization rotation measurement on a calibration sample, such as a sputtered gold film or GaAs, which exhibits no birefringence or Kerr effect.  We find that in a typical experiment we can preserve the linear polarization to better than a percent, but the setup birefringence is typically  $(\epsilon_1+\epsilon_2)\sim3-5\%$ (Fig.~\ref{fig:GaAs_test}).  However,  sample-induced birefringence might be much smaller than that, and therefore overwhelmed by the setup contribution. A considerable improvement can be obtained by using a modulation technique, whereby the sample response is modified by an external parameter (temperature, magnetic field).

\subsection{Temperature modulation}

Temperature modulation is achieved by focusing a second laser beam ($\unit[780]{nm}$ in our case), which we refer to as the source, on the same spot. The source beam is chopped at $\unit{kHz}$ frequencies, providing two key improvements over a non-modulated measurement: (1) the setup contribution is dramatically reduced and (2) measurements are performed using the lock-in technique at frequencies above $1/f$ noise, allowing for a measurement at the shot noise level.

Let the temperature dependent sample Jones matrix be: 

\begin{equation}
\begin{split}
\MoveEqLeft
J_{sam}\left(r,d_tr, d_tb,d_tk, \phi_0\right)=\\ 
&R\left(-\phi_{0}\right)\left(
\begin{array}{cc}
 r+d_tr+d_tb & d_tk \\
 -d_tk & r+d_tr-d_tb \\
\end{array}
\right)R\left(\phi_{0}\right),
\end{split}
\end{equation}
where $d_tr$, $d_tb$ and $d_tk$ denote temperature derivatives of reflectivity, birefringence, and the polar Kerr signal. 

\subsubsection{Ideal setup}

In the absence of the setup artifacts, the dominant signal proportional to the temperature modulation is the temperature derivative of Eq.~\ref{eq:noMod_ideal}. If there is no change of principal axes with temperature, the derivative reads: 

\begin{equation}\label{eq:thermalIdeal}
\begin{split}
\MoveEqLeft
   \frac{I\left(\phi\right)}{r^2}= \\ 
   & -2\left(\frac{d_tb}{r}+\frac{b}{r}\frac{d_tr}{r}\right) \sin (2 (\phi +\phi_0 ))\\
   & -2\left(\frac{d_tk}{r}+\frac{k}{r}\frac{d_tr}{r}\right).
\end{split}
\end{equation}
The birefringence and polar Kerr effect remain well separated as a sinusoidal variation and the offset. There are two contributions to each, one proportional to the temperature derivative of the quantity of interest ($d_tb$ and $d_tk$ for birefringence and Kerr, respectively), and the other one proportional to the product of the temperature derivative of reflectivity ($d_tr$) and the birefringence/polar Kerr. Unless $d_tr \gg d_tb, d_tk$, Eq.~\ref{eq:thermalIdeal} is dominated by the  thermal derivatives  $d_tb$ and $d_tk$, because $b/r\ll1$ and $k/r\ll1$, yielding: 

\begin{equation}\label{eq:thermalIdealApprox}
   \frac{I\left(\phi\right)}{r^2} \approx -2\frac{d_tb}{r}\sin (2 (\phi +\phi_0 ))-2\frac{d_tk}{r}.
\end{equation}
Although there is no reason to expect $d_tr \gg d_tb, d_tk$,  it is also not necessary to assume this: $d_tr$ can be directly measured in the same setup, by omitting the balanced measurement at the end and measuring the total signal, $I_{T} = |E_{V}|^2+|E_{H}|^2$, instead. In EuCd$_2$P$_2$ $d_tr$, $d_tb$ and $d_tk$ are all comparable, guaranteeing the validity of Eq.~\ref{eq:thermalIdealApprox}.

\subsubsection{Setup birefringence and retardance}
The first-order setup contribution to the thermally modulated measurement is proportional to the temperature derivative of Eq.~\ref{eq:setup}, whose only temperature dependent part is $r$, yielding: 

\begin{equation}\label{eq:setup_dt}
\begin{split}
\MoveEqLeft
   \frac{I_{setup}\left(\phi\right)}{r^2}=\\ 
   & 2\frac{d_tr}{r}\left((\epsilon_1+\epsilon_2)\sin(2\phi) -\sin^{2}\left(\frac{\delta_1+\delta_2}{2}\right)\sin(4\phi)\right).
\end{split}
\end{equation}
The size of this term can be directly estimated by measuring the setup response with no modulation (Fig.~\ref{fig:GaAs_test}), and multiplying it by the temperature derivative of reflectivity. In EuCd$_2$P$_2$ the setup contribution is at most $\sim 3-5\%$ of the real signal, rendering it insignificant. Furthermore, the artifacts described by Eq.~\ref{eq:setup} could never account for the observed domain structure because they do not depend on the principal axis orientation $\phi_0$.

\subsubsection{Cross terms}

The remaining question is whether there are any cross terms which could make a combination of real birefringence and setup artifacts appear like an offset, and vice versa. We find two such terms in our simulation:

\begin{enumerate}
    \item Finite birefringence, no polar Kerr effect: 
    \begin{equation}\label{eq:setupCT_b}
    \begin{split}
    \MoveEqLeft
       \frac{I_{CT1}\left(\phi\right)}{r^2}=\\ & \frac{d_tb}{r}\left(\cos(\delta_{2})\epsilon_{1}-\cos(\delta_{1})\epsilon_{2})\sin(2\phi_{0}\right).
    \end{split}
    \end{equation}
    A spurious offset can be observed in this case:  it depends on the domain orientation with respect to the laboratory coordinate system ($\sin\left(2\phi_0\right)$), and is modulated by the setup birefringence. This effect can therefore yield an offset which has a different value in each of the birefringent domains, is a fraction of the magnitude of the birefringence, and follows its temperature dependence. None of these apply to the offset we measure in EuCd$_2$P$_2$, confirming its origin as the polar Kerr effect.
    \item Finite polar Kerr effect, no birefringence\label{sec:crossTalk}:
    
    \begin{equation}\label{eq:setupCT_b}
       \frac{I_{CT2}\left(\phi\right)}{r^2}= \frac{2d_tk}{r}\epsilon_{1}\cos(\delta_{2})\cos(2\phi).
    \end{equation}
    The $\cos(2\phi)$ dependence can be in principle observed in this case, but it is again expected to be a fraction (proportional to setup birefringence) of the Kerr effect signal, inherit its temperature dependence, and its only spatial variation arises from the change of the sign of $d_tk$ when moving between ferromagnetic domains. None of this applies for the sinusoidal signal measured in thermal modulation in EuCd$_2$P$_2$, proving once again that the offset and sinusoidal variation are independent. 
\end{enumerate}

\subsection{Magnetic field modulation}
Much of the mathematical description outlined above applies equally well to the description of the field modulated experiment. However the relative size of the terms describing the sample response requires a modified discussion, and experimental setup.

\begin{figure}[b] 
\centering
\includegraphics[width=0.9\columnwidth]{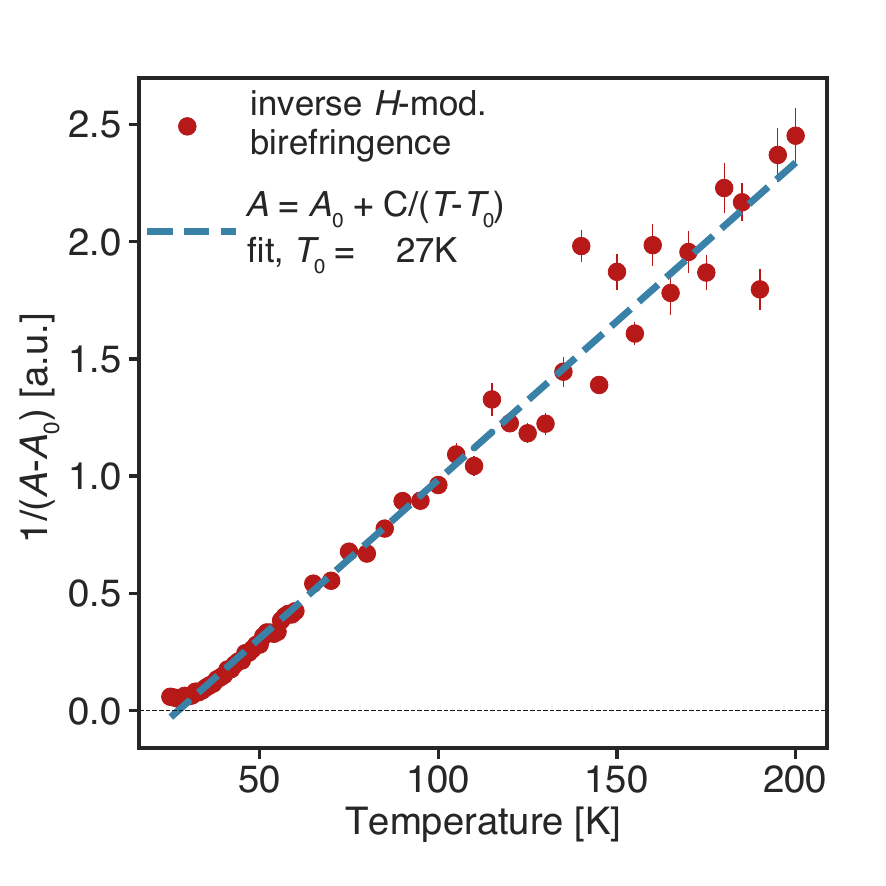}
\caption{Inverse field-modulated birefringence amplitude follows the Curie-Weiss law, indicating that it is in fact a measure of the artifact described by Eq.~\ref{eq:setupCT_chi}}.
\label{fig:susceptibility}
\end{figure}

As discussed in the
main text, sinusoidal polarization variation of the field-modulated signal is direct evidence of time-reversal symmetry breaking coexisting with $C_3$ symmetry breaking. However, the applied magnetic field introduces a magnetic moment proportional to the magnetic susceptibility. This magnetic moment induces a polar Kerr effect, and could in principle be detected as an offset in the field-modulated experiment (in practice this is complicated by the Faraday rotation of the cryostat windows). As the magnetic susceptibility increases with decreasing temperature, it is a real concern that the cross term described in Sec.~\ref{sec:crossTalk} (Eq.~\ref{eq:setupCT_b}) may prevent a clear determination of the onset of time-reversal symmetry breaking. In this context the cross artifact takes the form: 

\begin{equation}\label{eq:setupCT_chi}
       \frac{I_{CT2}\left(\phi\right)}{r^2}\sim \frac{\chi\left(T\right)}{r}\epsilon_{1}\cos(\delta_{2})\cos(2\phi),
    \end{equation}
with $\chi\left(T\right)$ denoting the magnetic susceptibility. Indeed, we observe this artifact. When we first attempted to measure the field-modulated birefringence, we observed a sinusoidal signal which persisted much above the magnetic transition temperature, and whose amplitude obeyed the Curie-Weiss law, with the same $T_{CW}=\unit[27]{K}$ as measured in the bulk susceptibility measurements \cite{wang_colossal_2021} (Fig.~\ref{fig:susceptibility}). Clearly, this was not a measure of a time-reversal symmetry-breaking order parameter, but of susceptibility through the cross-coupling term of the form given by Eq.~\ref{eq:setupCT_chi}. These measurements were performed with the same setup which yielded unambiguous separation of the polar Kerr effect and birefringence in the thermally modulated experiment, emphasizing the importance of carefully considering all possible sources of systematic errors in each experiment. 

\begin{figure}[t] 
\centering
\includegraphics[width=0.9\columnwidth]{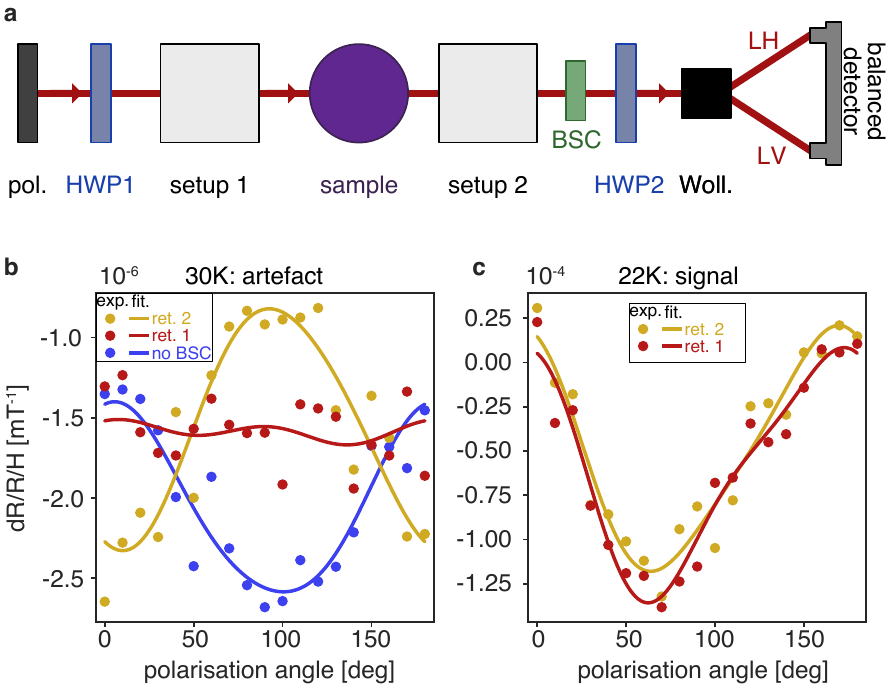}
\caption{(a) Simplified schematic of the setup including the Babinet–Soleil compensator (BSC), \emph{c.f.} Fig.~\ref{fig:schemeSimple}. (b) Field-modulated birefringence measured at \unit[30]{K}, without the BSC and for two different values of BSC retardance. The fact that the signal depends solely on the retardance is evidence that it is an artifact described by Eq.~\ref{eq:setupCT_chi}. (c) The real signal at $\unit[22]{K}$ is much larger than the artifact, and robust against retardance changes.}
\label{fig:SB_artifact}
\end{figure}

In order to mitigate this artifact, we note that it depends only on the retardance of the part of the setup between the sample and the detection ($\delta_{2}$), and the birefringence of the part of the setup before the sample $\epsilon_{1}$.  We therefore chose to modify the retardance of the second part of the setup by adding a continuously variable retarder (Babinet–Soleil compensator, BSC) after the sample (Fig.~\ref{fig:SB_artifact}a).

We observed that the sinusoidal signal at $\unit[30]{K}$ could be entirely changed by changing the retardance on the BSC (Fig.~\ref{fig:SB_artifact}b), proving that it is indeed caused by the cross term (Eq.~\ref{eq:setupCT_chi}). We therefore calibrated the setup by choosing a value of retardance which completely cancelled the artifact at $\unit[30]{K}$ (red in Fig.\ref{fig:SB_artifact}b). A thus calibrated setup was truly sensitive to time-reversal symmetry-breaking, and allowed to the observation of a sharp time reversal breaking transition at $T^{*}$. To demonstrate that this real signal is not significantly influenced by the choice of the BSC retardance, we measure the signal at $\unit[22]{K}$ at a few values of retardance; we observe no strong dependence (Fig.\ref{fig:SB_artifact}b). Upon comparison of Fig.\ref{fig:SB_artifact}(a,b), it is also clear that the real signal is much larger than the artifact. Nonetheless, observing, recognizing and mitigating the artifact was crucial in developing field-modulated birefringence as the sensitive probe of time-reversal symmetry-breaking that it has proven to be.

\section{Resonant elastic x-ray scattering}
\def \ecp{EuCd$_2$P$_2$}
\def \kafm{\textbf{k}$_{\textnormal{AFM}}$}
\def \kfm{\textbf{k}$_{\textnormal{FM}}$}

\begin{figure}[b]
	\includegraphics[width=0.9\columnwidth]{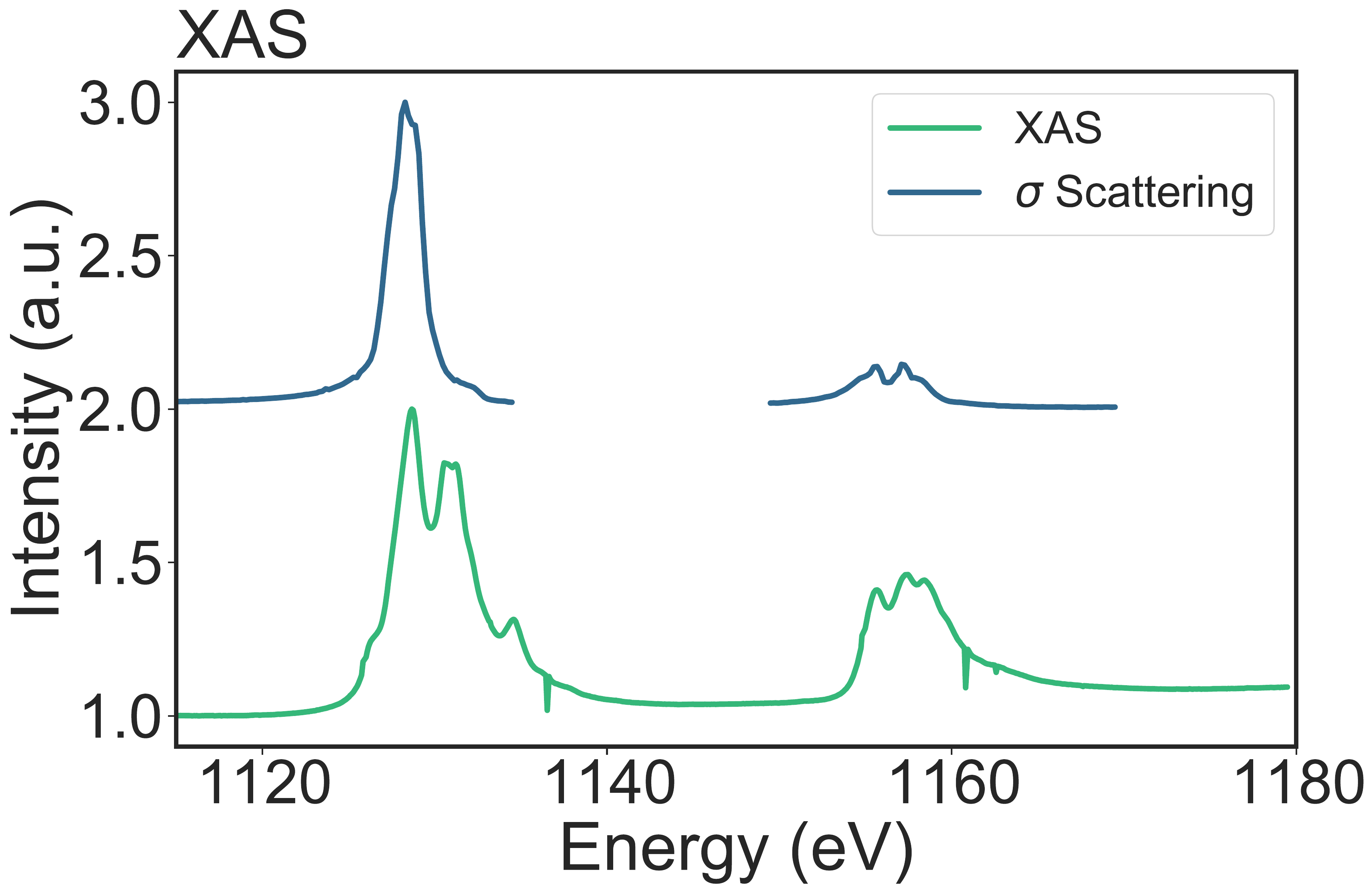}
    \caption{Energy-dependent scattering at (0 0 1/2) (top) and X-ray absorption (XAS) (bottom) of \ecp. Energy-dependent scattering shows that the AFM magnetic Bragg peak resonates at the Eu M5 edge.}
    \label{fig:xas}
\end{figure}

Resonant elastic X-ray scattering (REXS) is sensitive to magnetic ordering, making it an ideal technique to investigate the magnetic behavior of \ecp. To enhance the magnetic X-ray scattering of the Eu ions, we tuned the X-ray photon energy to the Eu $M_5$ edge (1127.5 eV), as verified by X-ray absorption spectroscopy (XAS) (Fig.~\ref{fig:xas}). The XAS measurements were conducted in total electron yield mode.

\begin{figure}[t]
	\includegraphics[width=0.85\columnwidth]{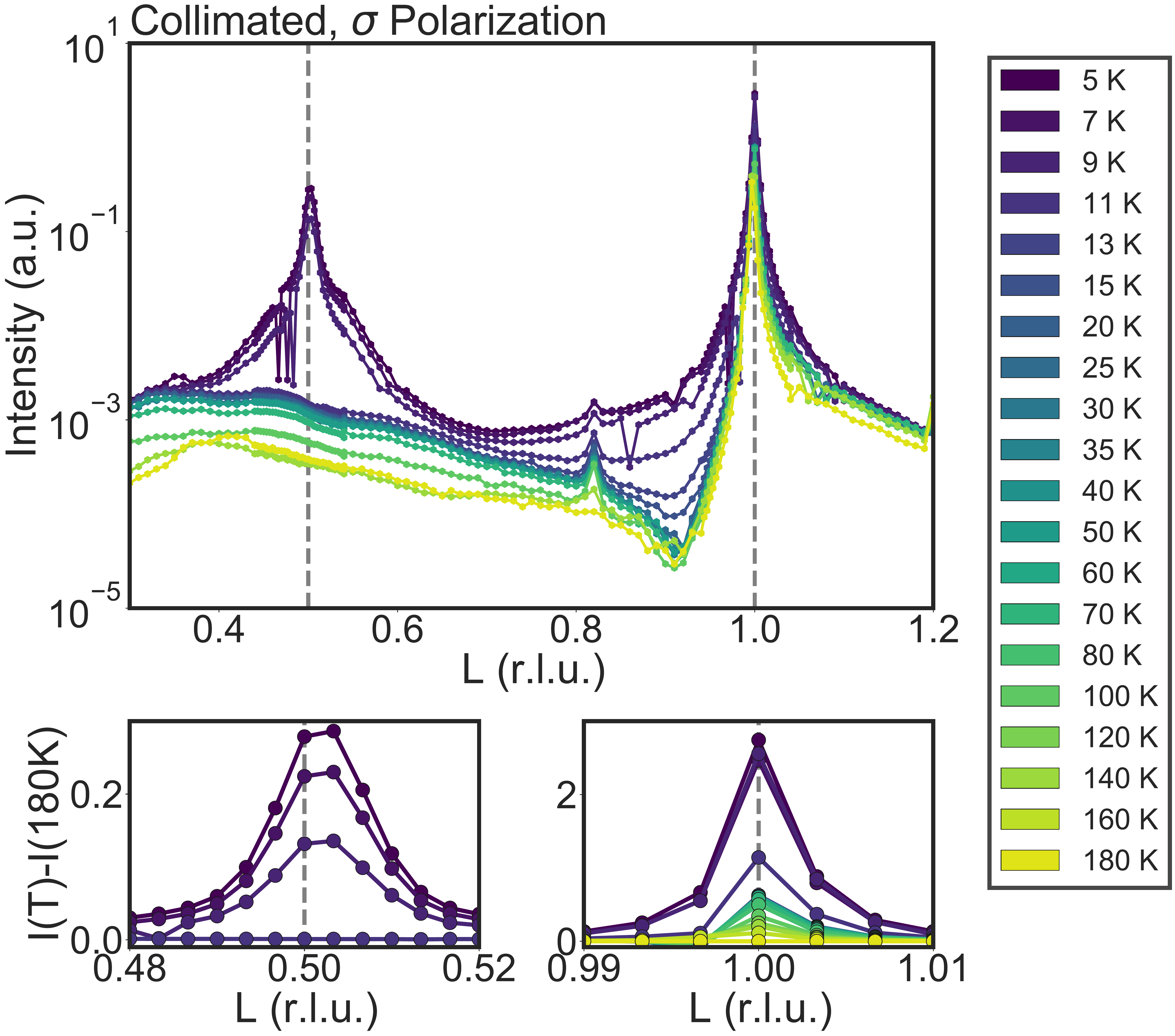}
    \caption{L-dependence of scattered X-ray intensity of $\sigma$-polarized light and a collimated beam. Lower panels show intensities with high-temperature contributions subtracted.}
    \label{fig:col_h}
\end{figure}

\begin{figure}[t]
	\includegraphics[width=0.85\columnwidth]{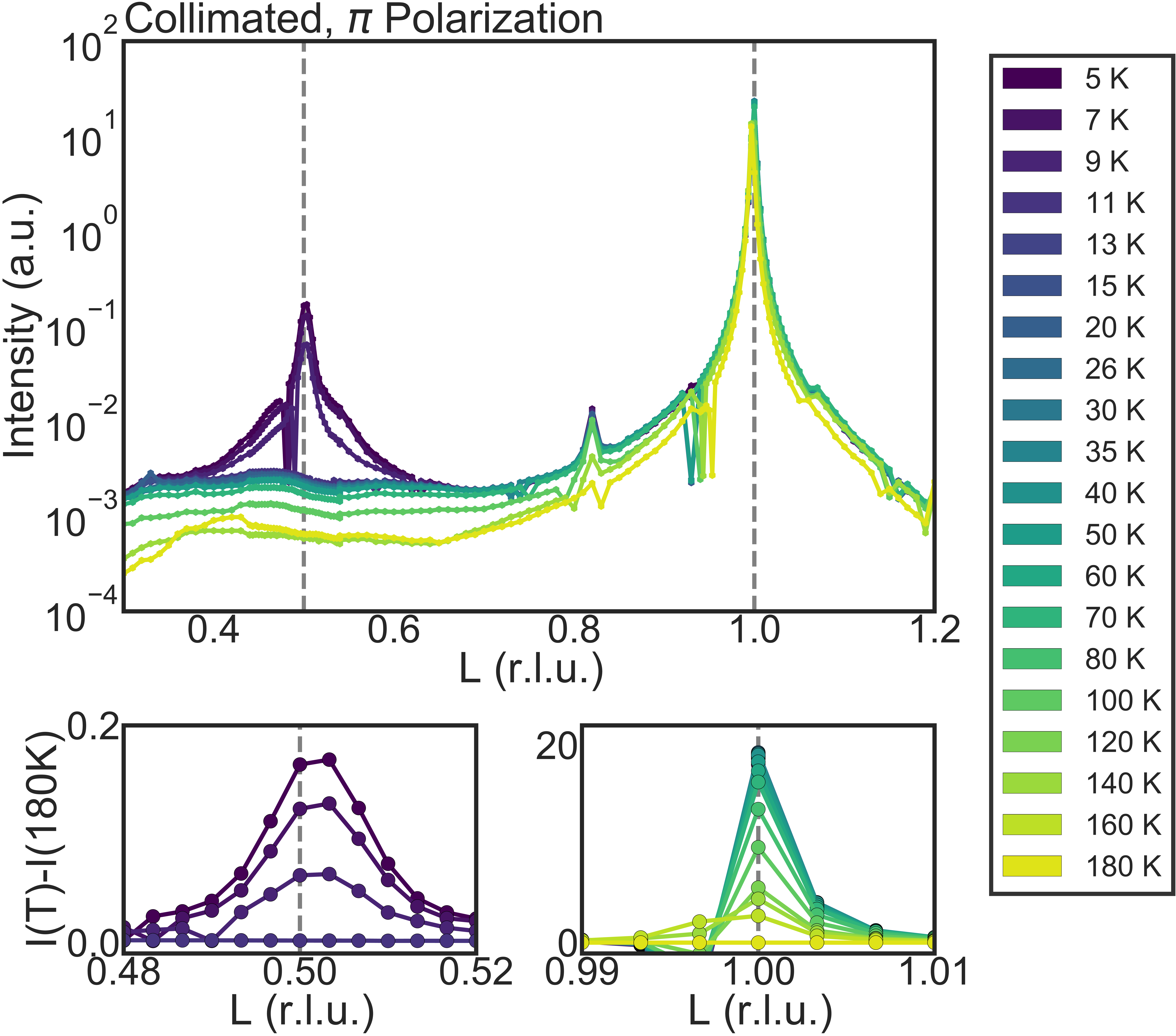}
    \caption{L-dependence of scattered X-ray intensity of $\pi$-polarized light and a collimated beam. Lower panels show intensities with high-temperature contributions subtracted.}
    \label{fig:col_v}
\end{figure}

A-type AFM ordering doubles the structural unit cell along the crystallographic c-axis, resulting in a magnetic Bragg peak at \kafm = (0 0 1/2). Contributions from ferromagnetic correlations at \kfm = (0 0 1), meanwhile, overlap with the structural Bragg peak. However, the scattering geometry provided a convenient workaround to study the FM behavior. The c-axis lattice parameter ($c=7.177$ \AA) means that the (1 0 0) reflection is near $2\theta=90 \deg$. This leads to a dip near $L=1$ ($L~\sim0.95$) due to Thomson scattering suppression (Fig.~\ref{fig:col_h}), which is absent with vertically ($\pi$) polarized X-rays (Fig.~\ref{fig:col_v}). The scattering contribution from the broad ferromagnetic Bragg peak fills in this dip, allowing us to measure the change of the FM order as a function of temperature.
The measurements shown in Fig.~2e of the main text were performed with $\sigma$-polarized light and a collimated beam of size 1.5mm $\times$ 0.4mm, therefore averaging the behavior over many domains seen in the optical measurements.

 Additionally, energy-dependent scattering was conducted by varying the energy of the incident photons and measuring the scattered intensity at the (0 0 1/2) antiferromagnetic q-vector. Energy-dependent scattering, when compared with the XAS, confirms that the AFM Bragg peak's resonant energy is at the Eu M5 edge (Fig.~\ref{fig:xas}).

\begin{figure}[t]
	\includegraphics[width=0.9\columnwidth]{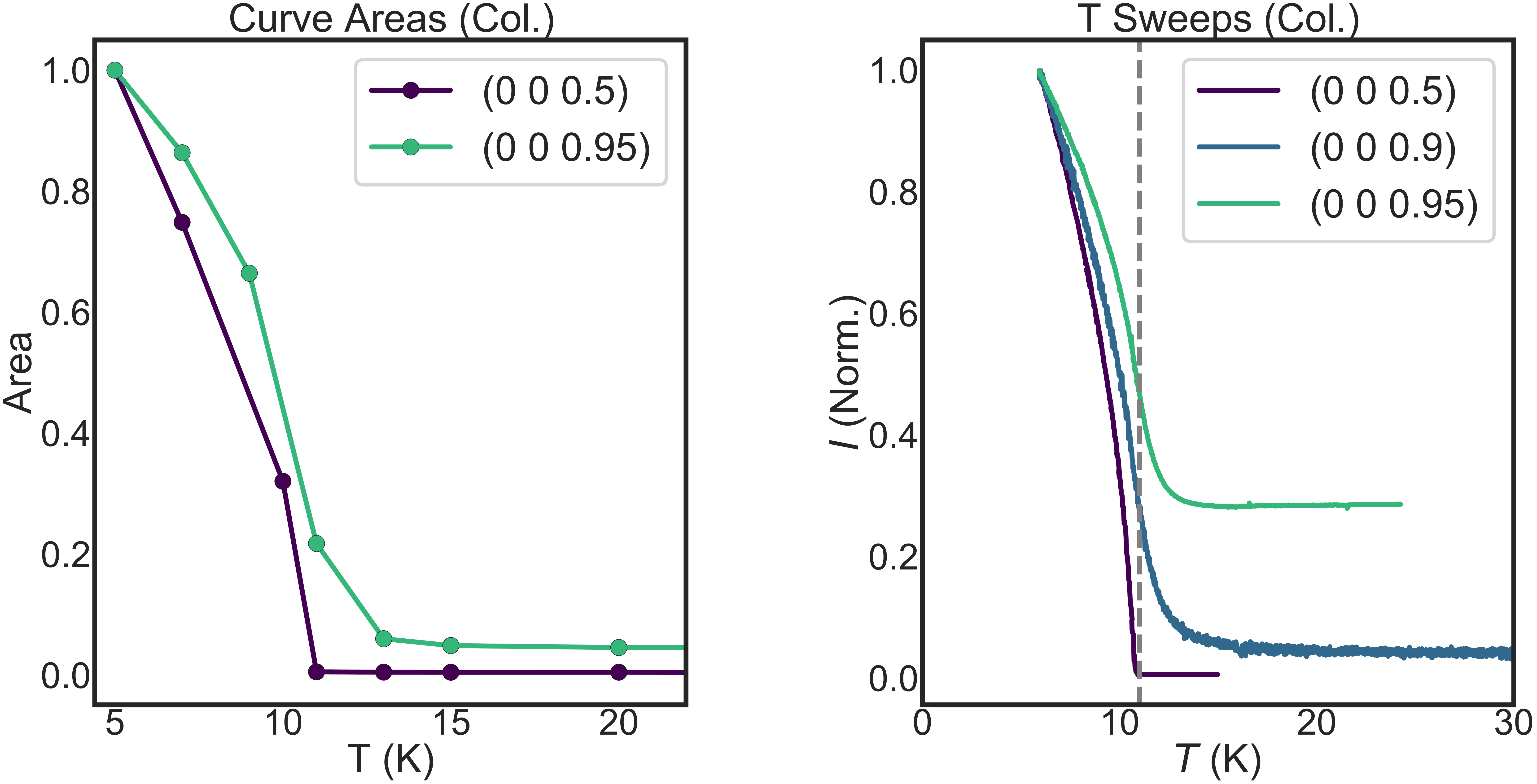}
    \caption{Analysis of L-dependence and rocking curves. The left panel shows the change in the rocking curve areas as a function of temperature, taken around (0 0 0.5) and (0 0 0.95). These are in agreement with the temperature sweeps shown in the right panel, and in Fig.~2e of the main text.}
    \label{fig:areas}
\end{figure}

Analysis of the temperature-dependence of the magnetic Bragg peaks is shown in Fig.~\ref{fig:areas} (right panel). A full temperature-dependence was acquired directly by aligning to the desired q-value and scanning over time while increasing and monitoring the sample temperature. The behavior of this measurement is in agreement with the behavior of integrated intensities extracted from the data in Fig.~\ref{fig:col_h} (Fig.~\ref{fig:areas}, left panel).  An order-parameter like transition is seen at $T_N=11$K in the AFM Bragg peak, in agreement with heat capacity and optical measurements. In contrast, the ferromagnetism has a smooth onset at $T_{FM}\textgreater\, T_N$, again consistent with the optical measurements. 

\section{Linear magneto-birefringence: symmetry considerations}

Onsager's relation states that  $\varepsilon_{ij}\left(S\right)=\varepsilon_{ji}\left(\Theta S\right)$, where $\varepsilon$ is the dielectric tensor,  $\Theta$ is the time-reversal operator and $S$ represents the system. Expanding the dielectric tensor to first order in applied magnetic field defines a third-rank linear magneto-optic tensor $\alpha_{ijk}$, such that $\varepsilon_{ij}(H_k)=\varepsilon_{ij}(0)+\alpha_{ijk}H_k$. According to Onsager reciprocity, $\alpha_{ijk}\left(S\right)=-\alpha_{jik}\left(\Theta S\right)$ and therefore systems that are time-reversal invariant must show $\alpha_{ijk}=-\alpha_{jik}$. It directly follows that time-reversal symmetry must be broken if any components of the linear magneto-optic tensor that are symmetric upon interchange of $ij$ are nonzero. Birefringence manifests through the symmetric part of the dielectric tensor ($\varepsilon_{ij}=\varepsilon_{ji}$), so nonzero linear magneto-birefringence (LMB) directly proves breaking of both time-reversal symmetry and $C_3$ symmetry.

\subsection{LMB terms allowed in EuCd$_2$P$_2$}

\begin{figure}[b] 
\centering
\includegraphics[width=0.9\columnwidth]{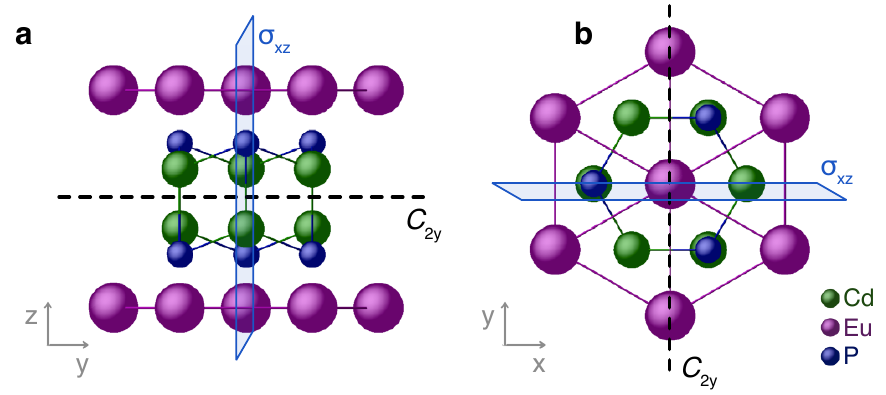}
\caption{Structure of EuCd$_2$P$_2$ viewed along the (a) $x$ and (b) $z$ axis, together with the rotational axis $C_{2y}$ and the mirror plane $\sigma_{xz}$.}
\label{fig:structure}
\end{figure}

Here we discuss how to determine which time-reversal symmetry-breaking order parameters allow which of the linear magneto-birefringence (LMB) terms ($\beta$, $\gamma$) in EuCd$_2$P$_2$. From the definition of the linear magneto-optic tensor $\alpha_{ijk}$, it follows that

\begin{equation}\label{eq:polarization}
   P_{i}=\alpha_{ijk}E_{j}H_k,
\end{equation}
where $P$ is the polarization and $E$ is the electric field. In order to find out which of the LMB terms ($\beta=\alpha_{xxz}-\alpha_{yyz}$, $\gamma=\alpha_{xyz}$) are symmetry-allowed, we impose that Eq.~\ref{eq:polarization} needs to be invariant under the point group symmetry operations. Since we know that the sample is birefringent, and therefore that the $C_{3z}$ symmetry is broken, it is sufficient to study the effect of the two-fold rotation $C_{2y}$ and the mirror plane $\sigma_{xz}$ (Fig.~\ref{fig:structure}).

Transformations of the components of the polar ($P_{i}$, $E_{j}$) and axial ($H_k$) vectors under those operations are shown in Table~\ref{tab:symmetries}, in which the $+$  indicates that the vector remains invariant under the transformation, and the $-$ indicates that it changes sign. The requirement of invariance of Eq.~\ref{eq:polarization} under the two symmetry operations determines the transformations of $\beta$ and $\gamma$ (Table~\ref{tab:symmetries}): $\beta$ is odd under both transformations, and $\gamma$ even. The final task is to identify the order parameters of the same symmetry as $\beta$ and $\gamma$.

 \begin{table}[h]
\centering
\begin{tabular}{|c |c| c| c|c| c|c||c|c|} 
 \hline
  & $x$ & $y$ & $z$ & $H_x$ & $H_y$ & $H_z$ & $\beta$ & $\gamma$ \\ 
 \hline\hline
 $C_{2y}$ & $-$ & $+$ & $-$ & $-$ & $+$ & $-$ &$-$ &$+$ \\ 
 \hline
 $\sigma_{xz}$ & $+$ & $-$ & $+$ & $-$ & $+$ & $-$&$-$ &$+$\\
 \hline
\end{tabular}
\caption{Transformation of the three orthogonal polar ($x$, $y$, $z$) and axial ($H_x$, $H_y$, $H_z$) vectors under the symmetry operations of the point group $2/m$.}
\label{tab:symmetries}
\end{table}

\subsection{Magnetic point groups and order parameters}
In Table~\ref{tab:OrderParameters}, we list the transformations of the four $C_3$-breaking order parameters (two magnetizations and two N\'{e}el vectors) under $C_{2y}$ and $\sigma_{xz}$, as well as the associated point groups. 

\begin{table}[h]
\centering
\begin{tabular}{| c| c|c| c|c|} 
 \hline
   & $M_x$ & $M_y$  & $L_x$ & $L_y$\\ 
 \hline\hline
 $C_{2y}$  &  - & + & + & - \\ 
 \hline
 $\sigma_{xz}$ & - & +  & - & +  \\ 
 \hline
 point group & $2'/m'$ & $2/m$ &  $2/m'$ & $2'/m$  \\ 
 \hline
\end{tabular}
\caption{Transformation of the $C_3$-breaking magnetic order parameters under the point group operations. The last row denotes the magnetic point groups that the order parameters belong to.}
\label{tab:OrderParameters}
\end{table}

A few conclusions can be drawn: 
\begin{itemize}
    \item Diagonal LMB term ($\beta$) has the same symmetry as $M_x$. Both are allowed in the magnetic point group $2'/m'$.
    \item Off-diagonal LMB term ($\gamma$) has the same symmetry as $M_y$. Both are allowed in the magnetic point group $2/m$.
    \item The AFM order parameters $L_x$ and $L_y$ belong to point groups which do not allow for LMB.
    \item Out-of-plane magnetic field ($H_z$) and magnetisation ($M_z$) transform in the same way as $M_x$. Therefore, coexisting $M_x$ and $M_z$ allow only for diagonal LMB ($\beta$).
    \item In contrast, a combination of $M_z$ and $M_y$ (or  $M_x$ and $M_y$) breaks all the symmetries; $\beta$ and $\gamma$ are then both allowed. 
\end{itemize}

These findings are summarized in Fig.~3b of the main text. 

\subsection{Any point group}

In general, one can ask whether a magnetic point group allows for either diagonal or off-diagonal LMB. The answer can be derived for every specific case, as demonstrated above, but it can also be found by consulting tabulated forms of symmetry allowed tensors for each magnetic point group \cite{birss_symmetry_1964}. The starting point is noting that the linear-magneto optic tensor $\alpha_{ijk}$ is a third-rank axial $c$-tensor, as defined by Birss\cite{birss_symmetry_1964}. The task is therefore to check in Table 7 of `Symmetry and Magnetism'\cite{birss_symmetry_1964} what form does a third rank axial $c$-tensor take in the magnetic point group of interest. 
If we do this for the groups discussed above, we easily find  that no such tensor is allowed for $2'/m$ or $2/m'$, while tensor $B_3$ is allowed in $2/m$ and  $C_3$ is allowed in $2'/m'$.  We then use Table 4 to see the forms of the two tensors, and find that  $C_3$ allows only diagonal terms, and  $B_3$ allows only off-diagonal terms, which is consistent with the analysis above. 

\section{Free energy model}
As discussed in the main text, we describe the coupled Eu-electron system with the free energy:

\begin{equation}\label{eq:freeEnergy} 
F=\alpha_2m^{2}+\beta_2M^{2}+\beta_4 M^{4}+JmM-Hm-HM,
\end{equation}
where $m$ and $M$ are the magnetizations of electrons and Eu, respectively, $J$ is the coupling between them, and $H_{z}$ an externally applied field.
To find the free energy minima, we require the derivatives of $F$ with respect to $M$ and $m$ to vanish, yielding the coupled equations:

\begin{equation} \label{eq:amplitudeCoil} 
\begin{split}
\MoveEqLeft
\frac{\partial F}{\partial M}=2M\left(\beta_2+2\beta_4 M^2+\frac{J}{2}m\right)-H=0, \\
& \frac{\partial F}{\partial m}=2\alpha_2m+JM-H=0,
\end{split}
\end{equation}
and resulting in Eq.~6 of the main text.

\section{Optical conductivity and the Hall effect}
\subsection{Carrier density}
\begin{figure}[b] 
\centering
\includegraphics[width=0.8\columnwidth]{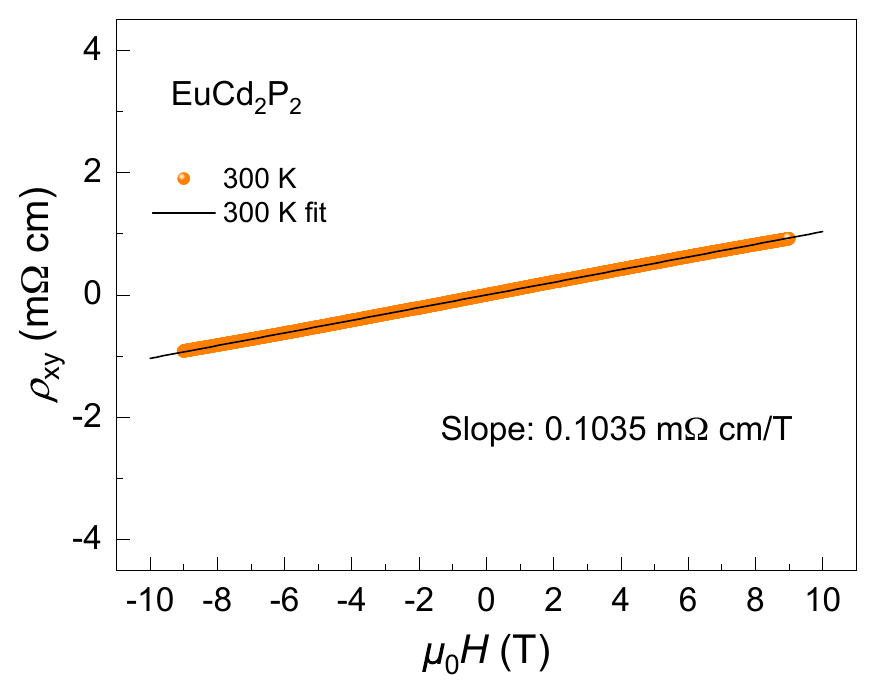}
\caption{The Hall resistivity as a function of magnetic field at  $\unit[300]{K}$.}
\label{fig:Hall}
\end{figure}

Room temperature Hall resistivity is linear in magnetic field (Fig.~\ref{fig:Hall}), and used to extract the carrier concentration using the usual relation: 

\begin{equation}
    n_e=\frac{1}{e\rho_{xy}}=\unit[0.6\times10^{19}]{cm^{-3}}.
\label{eq:Hall}
\end{equation}

\subsection{Mass and lifetime}
Within the Drude model, and in the absence of other excitations, the integrated spectral weight of conduction electrons is proportional to the ratio of carrier density and the effective mass: 

\begin{equation}
      \int_{0}^\infty \sigma(\omega)d\omega=\frac{\pi}{2}\frac{n_e e^2}{m}.
\end{equation}
Integrating the spectral weight of the peak in the inset of Fig.~5a of the main text, we find: 
\begin{equation}
      \frac{n_e}{m}= 0.4\times10^{19}\frac{\unit{cm^{-3}}}{m_e},
\end{equation}
where $m_e$ denotes the free electron mass. 
Combined with the carrier density extracted from the Hall effect (Eq.~\ref{eq:Hall}), we find: 
\begin{equation}
      m=1.5m_e.
\end{equation}
The scattering rate can be extracted from the ratio of the integrated spectral weight, and the zero frequency conductivity: 
\begin{equation}
    \Gamma=\frac{2}{\pi}\frac{\int_{0}^\infty \sigma(\omega)d\omega}{\sigma\left(\omega\right)}=\unit[9\times10^{13}]{s^{-1}}
\end{equation}

\bibstyle{apsrev4-2}

\bibliography{EuCd2P2_SI}